\documentclass[onecollarge,natbib]{svjour2}
\bibpunct{[}{]}{;}{n}{}{,} 
\smartqed  
\usepackage{natbib}
\usepackage{graphicx}
\usepackage{amssymb,amsmath}
\usepackage{comment}
\usepackage{mathptmx}      
\journalname{Few-Body Systems (EFB22)}
\begin{document}

\title{Fixed points of the Similarity Renormalization Group 
and the Nuclear Many-Body Problem~\thanks{
E.R.A. was supported by Spanish DGI (grant FIS2011-24149) and Junta de
Andaluc\'{\i}a (grant FQM225). S.S.  was supported by FAPESP
and V.S.T.  by FAEPEX and CNPq. Computational power
provided by FAPESP grants 2011/18211-2 and 2010/50646-6.
}
}


\author{E. Ruiz Arriola         \and
S. Szpigel \and V. S. Tim\'oteo
}


\institute{E. Ruiz Arriola \at
              Departamento
  de Fisica At\'omica, Molecular y Nuclear and Instituto Carlos I de
  Fisica Te\'orica y Computacional, Universidad de Granada, E-18071
  Granada, Spain.               
\email{earriola@ugr.es}           
\and S. Szpigel \at
              Faculdade de Computa\c c\~ao e Inform\'atica,
              Universidade Presbiteriana Mackenzie, S\~ao Paulo, Brazil.
              \email{szpigel@mackenzie.br} 
\and V. S. Tim\'oteo \at
              Grupo de \'Optica e Modelagem Num\'erica-GOMNI,
              Faculdade de Tecnologia, Universidade Estadual de
              Campinas-UNICAMP, Limeira, Brazil.  
\email{varese@ft.unicamp.br}
} \date{Presented by E. R. A. at 22th European Conference On Few-Body
  Problems In Physics: EFB22 \\ 9 - 13 Sep 2013, Krakow (Poland)}

\maketitle

\begin{abstract}
The Similarity Renormalization Group reduces the off-shellness
by driving the evolved interaction towards a diagonal band. We analyze
the infrared limit and the corresponding on-shell interactions and its
consequences for light nuclei.  Using a harmonic oscillator shell
model we obtain a Tjon line $B_\alpha= 4 B_t - 3 B_d$ which can be
understood from a combinatorics counting of nucleon pairs and
triplets in the triton and $\alpha$-particle and compares favorably
with realistic calculations.  
\end{abstract}

\def\bea{\begin{eqnarray}}
\def\eea{\end{eqnarray}}

\section{Introduction}
\label{intro}

Nuclear Physics has always been plagued with difficulties and
ambiguities related to off-shellness of two-body forces and the
inclusion of many-body forces deduced from the existence of equivalent
Hamiltonians~\cite{PhysRev.117.1590}.  In a remarkable paper Gl\"ockle
and Polyzou~\cite{polyzou1990three} point out that both problems are
actually intertwined since ``different off-shell extensions of
two-body forces can be equivalently realized as three-body
interactions'' and while ``there are no experiments measuring only
three-body binding energies and phase shifts that can determine if
there are no three-body forces in a three-body system'' it is still
likely that ``there may be some systems for which it is possible to
find a representation in which three-body forces are not needed''.

However, triton and $\alpha$-particle {\it ab initio} calculations
(see e.g.~\cite{Hammer:2012id} for a review) using 2-body high quality
interactions, i.e.  constrained to fit NN scattering data with $\chi^2
/{\rm d.o.f} \lesssim 1$ display exceedingly simple regularities, such
as the Tjon line, a linear correlation between the binding energies,
$B_\alpha = a B_t + b$ within a relatively wide range of energies (see
Fig.~\ref{fig:Tjon}). This suggests the onset of some scale
invariance~\cite{delfino2006few} and the understanding of this
correlation should also provide a credible value for the slope of the
Tjon line.

Actually, there is much freedom in making unitary transformations
keeping the two body bound and continuum spectrum while generating a
wide range of three- and four-body properties. A practical way to
generate these transformations is by means of the similarity
renormalization group (SRG) proposed by G\l azek and
Wilson~\cite{Glazek:1994qc} and independently by
Wegner~\cite{wegner1994flow}. The SRG was first used within Nuclear
Physics by Bogner, Furnstahl and Perry~\cite{Bogner:2006pc}. We
discuss how the above mentioned regularities could be understood
within a SRG context by analysing its infrared fixed points as we
discussed previously~\cite{Timoteo:2011tt}.  The proposed scenario is
free of ambiguites but indicates a predominant role played by 3-body
forces even in triton and alpha nuclei.

\section{SRG evolution and the fixed points}

\def\tr{{\rm Tr}}

The general SRG equation corresponds to a one-parameter 
operator evolution dynamics given by~\cite{Kehrein:2006ti},
\begin{eqnarray} 
\frac{d H_s}{ds} = [[ G_s, H_s],H_s]
\label{eq:SRG}
\end{eqnarray} 
and supplemented with an initial condition at $s=0$, $H_0$. With the
exception of some few
cases~\cite{szpigel2000quantum,Bogner:2007qb,Jones:2013cba} these
equations are mostly solved numerically~\cite{Szpigel:2010bj}, by
implementing a high momentum UV cut-off, $\Lambda$, and an infrared
momentum cut-off $\Delta p$, which reduces the analysis to the finite
$N$-dimensional case with $\Lambda = N \Delta p$. The isospectrality
of the SRG becomes evident from the trace invariance $\tr (H_s)^n= \tr
(H_0)^n$.

Fixed points of Eq.~(\ref{eq:SRG}) are given by stationary solutions,
$[[G_s,H_s],H_s]=0$ requiring $[G_s,H_s] = F(H_s)$. Thus there exists
a basis where both $G_s$ and $H_s$ can be simultaneously
block-diagonal for different energy subspaces with dimension equal the
degeneracy. The question is what choices of $G_s$ actually drive the
solution to this block-diagonal form. We will assume the usual
separation $H_0=T+V$. For generators which have the property $d/ds
(\tr G_s^2)=0$, and using cyclic properties of the trace and the
invariance of $\tr (H_s)^n$ one gets
\begin{eqnarray} \frac{d}{ds} \tr ( H_s-G_s)^2 = - 2 \tr (i[G_s,H_s])^\dagger
(i[G_s,H_s]) \le 0 \, .  \end{eqnarray} Because $\tr ( H_s-G_s)^2$ is positive
but its derivative is negative the limit for $s \to \infty$ exists and
corresponds to the fixed points, i.e. any starting $H_0$ gets indeed
diagonalized by the SRG equations~(\ref{eq:SRG}). Thus, the SRG
equation are just a continuous way of diagonalizing $H_0$.  The key
point is that there are many ways to diagonalize a Hamiltonian. For an
N-dimensional space there are $N!$ possible permutations regarding the
final {\it ordering} of states.  There are two complementary generator
choices: $G_s=H_D = {\rm diag} (H_s)$ by Wegner~\cite{wegner1994flow}
and by G\l azeck and Wilson~\cite{Glazek:1994qc} $G_s=T$.  More
generally $G_s=F(T)$~\cite{Li:2011sr}, describes a different
trajectory but identical fixed points. A choice in between is the
block-diagonal generator $G_s= P H_s P + Q H_s Q$ in two complementary
subspaces with $P+Q=1$~\cite{Anderson:2008mu,Arriola:2013era}. Denoting by $E_n$ the Hamiltonian and
$\epsilon_n$ the kinetic energy eigenvalues, the stability
analysis~\cite{Timoteo:2011tt} can be extended to show that ($C_{nm} $
is some suitable matrix encoding the initial $H_0$)

\begin{eqnarray}
H_{nm} (s)  = E_n \delta_{n,m}+ C_{nm}  
\begin{cases}
e^{-(\epsilon_n-\epsilon_m)(E_n-E_m) s} + \dots  \qquad {\rm Wilson}\\
e^{-(E_n-E_m)^2 s} + \dots \qquad \qquad \quad {\rm Wegner}
\end{cases}
\end{eqnarray} 
Therefore, while {\it all} $N!$ fixed points are stable in the Wegner
case, in the Wilson case there is a unique fixed point where the
Hamiltonian preserves the original ordering of states according to the
kinetic energy since $(\epsilon_n-\epsilon_m)(E_n-E_m) > 0$.  The
emerging crossing structure will be discussed in more detail in
Ref.~\cite{AST-prep}.

The previous considerations answer the question
of {\it how} similar are two given Hamiltonians in terms of the
Frobenius scalar product, norm and induced metric,
\begin{eqnarray} 
\langle A, B \rangle = {\rm Tr} (A^\dagger B) \, , \qquad 
|| V ||^2 \equiv {\rm Tr} (V^2)  \, , \qquad d(A,B) \equiv || A-B ||  \, . 
\end{eqnarray}
Moreover, because for $G_s=T$ the quantity 
$\tr (V_s^2)$ is non-negative and decreasing,  
we have an interesting variational property of the Wilson
generator, namely
\begin{eqnarray} 
 \lim_{s \to \infty} {\rm Tr} (V_s^2) =
  \min_{V} {\rm Tr} (V)^2 \Big|_{T+V = U H_0 U^\dagger}  \, .
\end{eqnarray}
Thus, the Wilson SRG drives the potential to the {\it smallest}, in
the Frobenius norm sense, possible one with the {\it same} spectrum.
In the continuum spectrum case, the orbital degeneracy induced by the
$d^3 p $ integration measure in the Frobenius norm has two
complementary effects: i) It suppresses low energy states and ii) it
enhances high energy components. Thus, minimizing $||V|| $ along the
SRG evolution transfers very efficiently high energy components into
small ones, and thus softens the possible cores. This is the main
reason why this approach became popular in Nuclear Physics where the
SRG cut-off $\lambda = 1/s^{\frac14}$ with energy dimensions is
used~\cite{Bogner:2006pc}.  In the many-body case where
$H=T+V_2+V_3+V_4 + \dots $ one subtracts CM motion, $G_s = T_{\rm rel}
\equiv T-T_{\rm CM}$ to preserve translational invariance and derive a
hierarchy of induced n-body SRG equations. Note that the SRG method
does not tell what the initial condition for the many body Hamiltonian
should be, and this amounts to switch-on many body forces at any value
of $\lambda$. Of course, once they are considered at any given
$\lambda$ the isospectrality guarantees that the n-body energy remains
independent of $\lambda$. This is checked for n=3 where the triton
binding energy is fixed with moderate bare three body forces,
$V_3(\lambda \to \infty)$~\cite{Jurgenson:2009qs,Jurgenson:2010wy,
  Hebeler:2012pr,Jurgenson:2013yya,Wendt:2013bla,Furnstahl:2013oba,Roth:2013fqa}.


Driving the potential to its diagonal form has the further advantage
of reducing calculations to first order perturbation theory and also
to get rid of the off-shell effects completely. For instance, the
$R$-matrix on the momentum grid $p_n$ reduces to the {\it on-shell
  potential} of the Lippmann-Schwinger equation; for the $^1S_0$ state
$\lim_{\lambda \to 0} V_{\lambda} (p_n,p_n) = - \tan \delta^{\rm LS}
(p_n)/p_n$ is a stable fixed point~\cite{Timoteo:2011tt}. We have
checked these trends regarding the fixed points numerically. Here, we
illustrate our points using the Wilson generator although many
features are common to the Wegner generator. Further details will be
presented in a forthcoming publication~\cite{AST-prep}.

\begin{figure}
\begin{center}
\includegraphics[height=6cm,width=6cm]{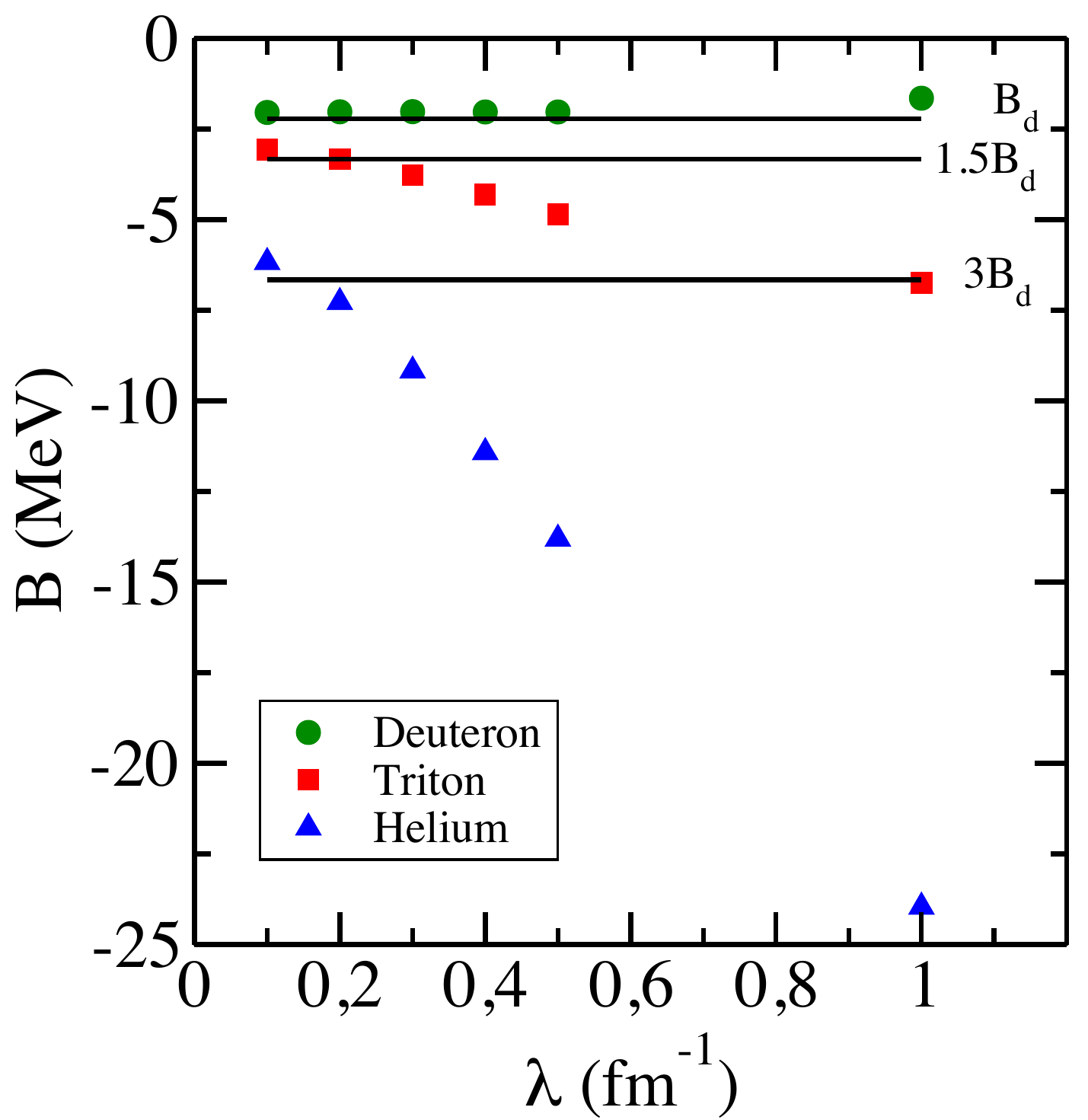}  
\hspace{0.5cm}
\includegraphics[height=6cm,width=6cm]{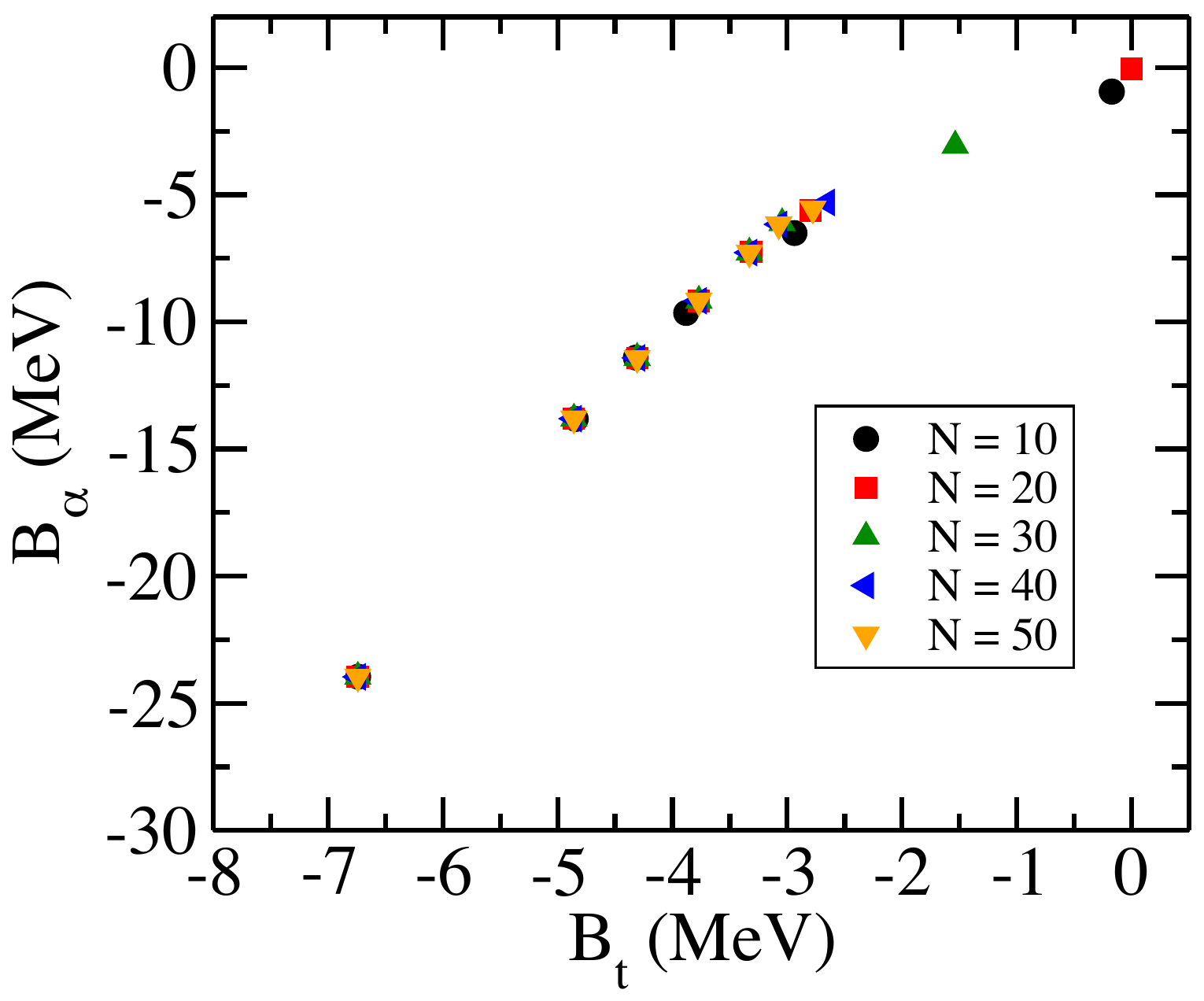} 
\end{center}
\caption{Variational Shell Model Binding Energies (in MeV) for
  different values of the SRG cut-off compared to the limiting
  on-shell values for $\lambda \to 0$ (left panel). $B_\alpha-B_t$
  Tjon line for different N values of the momentum grid (right
  panel). In both cases we assume 3- and 4-body forces to vanish at
  that scale $\lambda$, $V_3(\lambda)=V_4(\lambda)=0$. The 2-body potentials $V_2(\lambda)$ are phase equivalent.}
\label{fig:binding}
\end{figure}

\section{On-shell analysis of 2- , 3- and 4-body problem} 

In a recent work~\cite{Arriola:2013nja} we show how using simple
shell model variational calculations the binding energies of $^3$H,
$^4$He, $^{16}$O and $^{40}$Ca nuclei can be reasonably well described
by keeping the SRG evolved down to $\lambda \sim 1-2 {\rm fm}^{-1}$
two-body high quality potentials, i.e. interactions providing a
$\chi^2/{\rm d.o.f}\lesssim 1$. In order to keep the numerical effort
to a minimum and pursue the on-shell limit, $\lambda \to 0$, we use the
gaussian separable toy model of Ref.~\cite{Arriola:2013era} which
describes the $^1S_0$ and $^3S_1$ channels reasonably well. Within
such a scheme the energies are computed using  $|t \rangle \equiv 
|(1s)^3 \rangle $ and  $|\alpha \rangle \equiv 
|(1s)^4 \rangle $ with $1s$ HO wave functions with a $b$ parameter, 
\begin{eqnarray} 
\{ E_t(\lambda), E_\alpha(\lambda)\}=\{-B_t,-B_\alpha\} = \min_b \left[
  (A-1) \langle \frac{p^2}{2M} \rangle_{1s} + \frac{A(A-1)}{2} \frac12
  \langle V_{^1S_0,\lambda} + V_{^3 S_1,\lambda} \rangle_{{\rm rel},1s} \right]
\Big|_{A=3,4} 
\label{eq:b34}
\end{eqnarray} 
which can be interpreted in terms of the number of pairs in the
$^1S_0$ and $^3S_1$ states being $n_{^1S_0,t}=n_{^3S_1,t}=3/2$ for
triton and $n_{^1S_0 ,\alpha}=n_{^3S_1,\alpha}=6/2 $ for
$^4$He. Within this shell model scheme $E_d(\lambda)= \min_b \langle
p^2/M + V_{^3 S_1,\lambda} \rangle_{{\rm rel},1s}$. For $\lambda=\infty$
the deuteron is unbound by $0.2 {\rm MeV}$, and the binding energy
obtained for the triton is $B_t^{\rm Var}=5.9661 {\rm MeV}$ to be
compared with the exact Faddeev equation result $B_t=6.65543 {\rm
  MeV}$ whereas the $\alpha$-particle yields $B_\alpha^{\rm
  Var}=32.1054 {\rm MeV}$.

\begin{figure*}
\centering
  \includegraphics[width=0.7\textwidth]{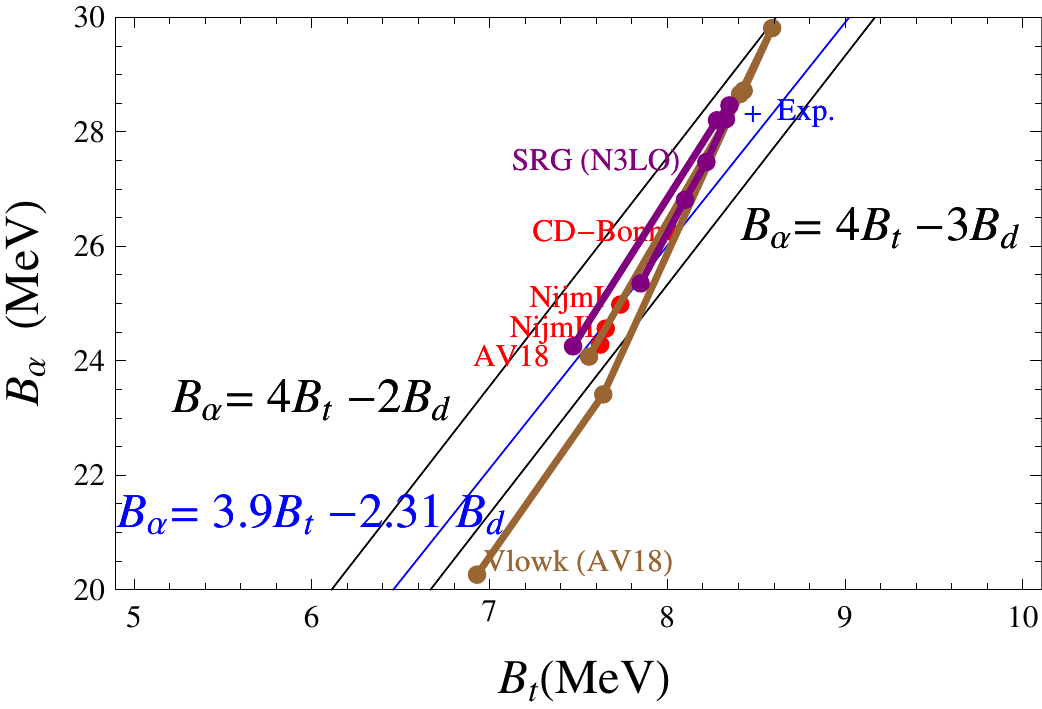}
\caption{Tjon line predicted by the on-shell 2-body interaction
  subjected to a 3-body force fulfilling $\langle V_3 \rangle_\alpha =
  4 \langle V_3 \rangle_t $ but assuming different number of
  deuterons. We compare with few body calculations {\it without}
  3-body forces~\cite{Hammer:2012id} and 2-body high quality
  potentials fitting NN scattering data and the deuteron.}
\label{fig:Tjon}       
\end{figure*}

It is interesting to analyze the SRG evolution from these
$\lambda=\infty$ values down to $\lambda=0$ {\it without} 3-body or
4-body forces, i.e. taking $V_3(\lambda)=V_4(\lambda)=0$. One can
show~\cite{AST-prep} that for the variational shell model 
\begin{eqnarray}
\lim_{\lambda \to 0 } E_d(\lambda) = -B_d \, ,\qquad
\lim_{\lambda \to 0 } E_t(\lambda) = -\frac32 B_d \, , \qquad
  \lim_{\lambda \to 0 } E_\alpha (\lambda) = - 3 B_d \, , 
\end{eqnarray}
which is checked by the numerical calculation (see
Fig.~\ref{fig:binding}).  We also find linear correlations in two
regimes
\begin{eqnarray}
\Delta B_\alpha / \Delta B_t \sim 2 \, (\lambda \to 0) \, , \qquad 
\Delta B_\alpha / \Delta B_t \sim 4 \, (\lambda \sim 1) \, .  
\end{eqnarray}
If we now switch on the SRG induced 3-body and 4-body forces, the
difference of the on-shell result to the original value corresponds to
the off-shellnes of the $\lambda=\infty$ two body potential,
\begin{eqnarray} -B_t = -\frac{3}{2} B_d+
\langle t | V_3 | t \rangle \, , \qquad
-B_\alpha = -3B_d 
+ \langle
  \alpha |V_3 + V_4| \alpha \rangle \, . 
\end{eqnarray} 
Neglecting $\langle \alpha |V_4 | \alpha \rangle $ and 
taking $\langle \alpha |V_3 | \alpha \rangle = 4 \langle t |V_3 | t
\rangle $ corresponding to 4 triplets in the $\alpha$-particle (see also~\cite{Sato01021974}) we get $B_\alpha = 4
B_t -3 B_d $ which gives $ B_\alpha = 4 \times 8.482 - 3 \times 2.225
= 27.53 \, ({\rm exp.}  28.296) \, {\rm MeV} $. Of course, one may say
the $^3$H =pnn contain 2 deuterons and $^4$He=ppnn contains 4
deuterons, so that $B_\alpha= 4B_t - 2B_d$.  In general, we may write  $n_{d/\alpha} = 2 n_{^3S_1,\alpha}$ and $n_{d/t} = 2 n_{^3S_1,t}$ and hence 
\begin{eqnarray} B_\alpha =
n_{t/\alpha} B_t + (n_{d/\alpha} - n_{d/t} n_{t/\alpha})B_d/2 \, . 
\label{eq:pair-triplet}
\end{eqnarray} 
Most calculations in Nuclear Physics use a mean field reference state
(often HO shell model) upon which correlations are built. For
instance, in Ref.~\cite{Feldmeier:2011qy} it is found that the number
of {\it correlated pairs} is given by
$n_{^1S_0,t}=1.490$,$n_{^3S_1,t}=1.361$, $n_{^1S_0,\alpha}=2.572$ and
$n_{^3S_1,\alpha}=2.992$ with $(B_d,B_t,B_\alpha )= (2.24,7.76,25.09)
{\rm MeV}$ which, from our Eq.~(\ref{eq:pair-triplet}), requires
$n_{t/\alpha}=3.9$ and then $B_\alpha= 3.9 B_t - 2.31 B_d$. These Tjon
lines are plotted in Fig.~\ref{fig:Tjon} using the different choices
in Eq.~(\ref{eq:pair-triplet}) and compared with several accurate
calculations using realistic 2-body forces (without 3- or 4-body
forces)~\cite{Hammer:2012id}.  In our interpretation, the slope of the
Tjon line is just the number of {\it correlated triplets} in $^4$He,
$(\partial B_\alpha/ \partial B_t)_{B_d}=n_{t/\alpha}=\langle V_3
\rangle_\alpha/ \langle V_3 \rangle_t$ for {\it on-shell}
interactions. These intriguing results suggest that $\langle \alpha |
V_4 | \alpha \rangle \lesssim 1 {\rm MeV}$ and will be analysed in
more detail elsewhere~\cite{AST-prep}.


\section{Conclusions}

SRG methods allow to reduce the two-body off-shell ambiguity {\it
  completely} when the infrared limit is taken and thus only {\it
  measurable} two-body information is needed. The same observation
applies to multi-body interactions, and while current calculations fix
the three body force from the triton binding energy, more work is
required to pin down how much three-body measurable input would
actually be demanded within such an on-shell scheme.  However, using
these ideas a simple explanation of the observed linear Tjon
correlations between the $\alpha$-particle and triton binding energies
emerges from purely combinatorial arguments. In this scheme the SRG
evolved 3-body forces are large in the infrared limit whereas 4-body
forces remain moderate.


\begin{thebibliography}{10}
\providecommand{\url}[1]{{#1}}
\providecommand{\urlprefix}{URL }
\expandafter\ifx\csname urlstyle\endcsname\relax
  \providecommand{\doi}[1]{DOI \discretionary{}{}{}#1}\else
  \providecommand{\doi}{DOI \discretionary{}{}{}\begingroup
  \urlstyle{rm}\Url}\fi

\bibitem{PhysRev.117.1590}
H.~Ekstein: 
Equivalent Hamiltonians in scattering theory.  
Phys. Rev. \textbf{117}, 1590 (1960)

\bibitem{polyzou1990three}
W.~Polyzou, W.~Gl{\"o}ckle: Three-body interactions and on-shell equivalent two-body interactions. Few-Body Systems \textbf{9}(2), 97 (1990)

\bibitem{Hammer:2012id}
H.W. Hammer, A.~Nogga, A.~Schwenk:
	Three-body forces: From cold atoms to nuclei. 
 Rev.Mod.Phys. \textbf{85}, 197 (2013)

\bibitem{delfino2006few}
A.~Delfino, T.~Frederico, V.~Tim{\'o}teo, L.~Tomio:
The few scales of nuclei and nuclear matter. Physics Letters B
  \textbf{634}(2), 185 (2006)

\bibitem{Glazek:1994qc}
S.D. Glazek, K.G. Wilson: 
	Perturbative renormalization group for Hamiltonians. Phys. Rev. \textbf{D49}, 4214 (1994)

\bibitem{wegner1994flow}
F.~Wegner: 
Flow equations for Hamiltonians. 
 Annalen der physik \textbf{506}(2), 77 (1994)

\bibitem{Bogner:2006pc}
S.~Bogner, R.~Furnstahl, R.~Perry:
	Similarity Renormalization Group for Nucleon-Nucleon Interactions. 
 Phys.Rev. \textbf{C75}, 061001 (2006)

\bibitem{Timoteo:2011tt}
V.~Timoteo, S.~Szpigel, E.~Ruiz~Arriola: 
	Symmetries of the Similarity Renormalization Group for Nuclear Forces. Phys.Rev. \textbf{C86}, 034002 (2011) 

\bibitem{Kehrein:2006ti}
S.~Kehrein, \emph{{The flow equation approach to many-particle systems}}
  (Springer, 2006)

\bibitem{szpigel2000quantum}
S.~Szpigel, R.~Perry, A.~Mitra, \emph{{Quantum Field Theory, A 20th Century
  Profile}} (Hindustan Publishing Co., 2000)

\bibitem{Bogner:2007qb}
S.~Bogner, R.~Furnstahl, R.~Perry:
	Three-Body Forces Produced by a Similarity Renormalization Group Transformation in a Simple Model.  Annals Phys. \textbf{323}, 1478 (2007)

\bibitem{Jones:2013cba}
B.D. Jones, R.J. Perry:
Similarity flow of a neutral scalar coupled to a fixed source. 
 ArXiv:nucl-th/1305.6599  (2013)

\bibitem{Szpigel:2010bj}
S.~Szpigel, V.S. Timoteo, F.d.O. Duraes:
	Similarity Renormalization Group Evolution of Chiral Effective Nucleon-Nucleon Potentials in the Subtracted Kernel Method Approach. 
 Annals Phys. \textbf{326}, 364 (2010)

\bibitem{Li:2011sr}
W.~Li, E.~Anderson, R.~Furnstahl:
	The Similarity Renormalization Group with Novel Generators. Phys.Rev. \textbf{C84}, 054002 (2011)

\bibitem{Anderson:2008mu}
E.~Anderson, S.~Bogner, R.~Furnstahl, E.~Jurgenson, R.~Perry, et~al.:
	Block Diagonalization using SRG Flow Equations. Phys.Rev.
  \textbf{C77}, 037001 (2008)

\bibitem{Arriola:2013era}
E.~Ruiz~Arriola, S.~Szpigel, V.~Timoteo: 
	Implicit vs Explicit Renormalization and Effective Interactions.  
Phys. Lett. \textbf{B 728}, 596 (2014).  

\bibitem{AST-prep}
E.~Ruiz~Arriola, S.~Szpigel, V.~Timoteo, In preparation  (2013)


\bibitem{Jurgenson:2009qs}
  E.~D.~Jurgenson, P.~Navratil and R.~J.~Furnstahl: 
  Evolution of Nuclear Many-Body Forces with the Similarity Renormalization Group.   Phys.\ Rev.\ Lett.\  {\bf 103} (2009) 082501


\bibitem{Jurgenson:2010wy} 
  E.~D.~Jurgenson, P.~Navratil and R.~J.~Furnstahl: 
Evolving Nuclear Many-Body Forces with the Similarity Renormalization Group,
  Phys.\ Rev.\ C {\bf 83}, 034301 (2011)


\bibitem{Hebeler:2012pr} 
  K.~Hebeler: Momentum space evolution of chiral three-nucleon forces. 
  Phys.\ Rev.\ C {\bf 85}, 021002 (2012)


\bibitem{Jurgenson:2013yya} 
  E.~D.~Jurgenson, P.~Maris, R.~J.~Furnstahl, P.~Navratil, W.~E.~Ormand and J.~P.~Vary: P-shell nuclei using Similarity Renormalization Group evolved three-nucleon interactions,
  Phys.\ Rev.\ C {\bf 87}, 054312 (2013)

\bibitem{Wendt:2013bla} 
  K.~A.~Wendt: 
  Similarity Renormalization Group Evolution of Three-Nucleon Forces in a Hyperspherical Momentum Representation. 
  Phys.\ Rev.\ C {\bf 87}, 061001 (2013)


\bibitem{Furnstahl:2013oba}
  R.~J.~Furnstahl and K.~Hebeler: 
  New applications of renormalization group methods in nuclear physics,''
  Rept.\ Prog.\ Phys.\  {\bf 76} (2013) 126301

\bibitem{Roth:2013fqa} 
  R.~Roth, A.~Calci, J.~Langhammer and S.~Binder: 
 Evolved Chiral NN+3N Hamiltonians for Ab Initio Nuclear Structure Calculations,
  arXiv:1311.3563 [nucl-th].




\bibitem{Arriola:2013nja}
E.~Ruiz~Arriola, V.~Timoteo, S.~Szpigel:
Nuclear Symmetries of the similarity renormalization group for nuclear forces. 
 PoS \textbf{CD12}, 106 (2013)

\bibitem{Sato01021974}
M.~Sato, Y.~Akaishi, H.~Tanaka:
Effects of Three-Body Force in the Triton and the Alpha Particle. Prog. of Theor. Phys. Supplement, \textbf{56}, 76 (1974) 

\bibitem{Feldmeier:2011qy}
H.~Feldmeier, W.~Horiuchi, T.~Neff, Y.~Suzuki:
	Universality of short-range nucleon-nucleon correlations.  
 Phys.Rev. \textbf{C84}, 054003 (2011)

\end{thebibliography}

\end{document}